\documentclass[
 aps,
 amsmath,amssymb,
 reprint,
 superscriptaddress,
 nofootinbib,
 floatfix,
 preprintnumbers
]{revtex4-1}
\usepackage[english]{babel}

\usepackage[letterpaper,top=2cm,bottom=2cm,left=3cm,right=3cm,marginparwidth=1.75cm]{geometry}

\usepackage[final]{changes}
\usepackage{amsmath}
\usepackage{bbold}
\usepackage{graphicx}
\usepackage{float}
\usepackage{braket}
\usepackage[colorlinks=true, allcolors=blue]{hyperref}
\usepackage[caption=false,subrefformat=parens,labelformat=parens]{subfig}

\newcommand{\eq}[1]{Eq.~\eqref{#1}}
\newcommand{\fig}[1]{Fig.~\ref{#1}}

\begin{document}

\title{
How many quantum gates do gauge theories require?
}

\author{Edison M.\ Murairi}
\email{emm712@gwu.edu}
\affiliation{Department of Physics, The George Washington University,
Washington, District of Columbia 20052, USA
}

\author{Michael J.\ Cervia}
\email{cervia@gwu.edu}
\affiliation{Department of Physics, The George Washington University,
Washington, District of Columbia 20052, USA
}
\affiliation{Department of Physics, University of Maryland,
College Park, Maryland 20742, USA
}

\author{Hersh Kumar}
\email{hekumar@umd.terpmail.edu}
\affiliation{Department of Physics, University of Maryland,
College Park, Maryland 20742, USA
}

\author{Paulo F.\ Bedaque}
\email{bedaque@umd.edu}
\affiliation{Department of Physics, University of Maryland,
College Park, Maryland 20742, USA
}

\author{Andrei Alexandru}
\email{aalexan@gwu.edu}
\affiliation{Department of Physics, The George Washington University,
Washington, District of Columbia 20052, USA
}
\affiliation{Department of Physics, University of Maryland,
College Park, Maryland 20742, USA
}

\date{\today}

\begin{abstract}
We discuss the implementation of lattice gauge theories on digital quantum computers, focusing primarily on the number of quantum gates required to simulate their time evolution.
We find that to compile quantum circuits, using available state-of-the-art methods with our own augmentations, the cost of a single time step of an elementary plaquette is beyond what is reasonably practical in the current era of quantum hardware.
However, we observe that such costs are highly sensitive to the truncation scheme used to derive different Hamiltonian formulations of non-Abelian gauge theories, emphasizing the need for low-dimensional truncations of such models in the same universality class as the desired theories.
\end{abstract}

\maketitle

\section{Introduction}

Among the many future applications of quantum computing to solving physics problems, the simulation of non-Abelian gauge theories is one of the most desired. Gauge theories describe all fundamental forces of the Universe, but many of their features are inherently non-perturbative and resistant to analytical analysis. Even stochastic numerical methods, despite recent advancements, leave almost untouched the many-body baryon problem and real-time observables, both of which are of great phenomenological importance.

Gauge theories, like any field theory, have an infinite-dimensional Hilbert space, while
digital quantum computers have finite and---for the foreseeable future---small quantum registers.
Therefore, some kind of truncation of the Hilbert space is necessary. One common truncation is the substitution of the continuum, infinite space by a finite lattice.
The use of a spatial lattice has the added benefit of providing an ultraviolet cutoff necessary for the non-perturbative definition of the theory.
Since this step has been extensively studied in the context of lattice gauge theories and their numerical simulation with (classical) computers, much is known about it.
For instance, the concept of universality controls the continuum limit that is necessary as the last step of these calculations.
It is found that a variety of lattice Hamiltonians lead to the same continuum limit and that some symmetries are essential in recovering the proper continuum limit (for instance, gauge symmetry), while others arise naturally in the continuum limit (for instance, rotational invariance).
In purely fermionic theories, the use of a finite spatial lattice is enough to render the Hilbert space finite-dimensional.
In contrast, bosonic theories, defined over even a single spatial point (equivalent to the quantum mechanics of a particle in one dimension), already have an infinite-dimensional Hilbert space.
As such, it is necessary to have a truncation of not only physical space but also {\it field} space. 

Several methods to accomplish this goal have been suggested in different models
~\cite{Mathur:2004kr,Byrnes:2005qx,Zohar:2012xf,Zohar:2013zla,Zohar:2014qma,Wiese:2014rla,Zohar:2016iic,Bender:2018rdp,Alexandru:2019nsa,Klco:2019evd,Raychowdhury:2019iki,Davoudi:2020yln,Kreshchuk:2020dla,Ciavarella:2021nmj,Alexandru:2021jpm}.
For some of these proposals, besides the spatial continuum limit, an independent field space continuum limit is required.
In this case, the dimension of the Hilbert space grows as the last limit is taken, and the issue we consider in the present paper (the cost in terms of quantum gates) becomes more severe.
In other proposals, an argument is made that, even with the field space truncated, the resulting theories \replaced{are}{is} in the same universality class of the targeted continuum theory.
In that case, the limit where dimension of the Hilbert space grows does not need to be explored, offering a great advantage in terms of simplicity.
The combined discretization of physical and field space is sometimes called {\it qubitization}.
For gauge theories in four spacetime dimensions, no general qubitization method has been established, although qubitizations have been found for simpler models \cite{Alexandru:2019ozf,Singh:2019uwd,Bhattacharya:2020gpm,Alexandru:2021xkf,Singh:2019jog}.
Nevertheless, there are some old discretized models that could play this role and have the proper symmetries and fairly small Hilbert spaces.
We use some of these models to estimate the gate-complexity of simulating gauge theories.

In Sec.~\ref{sec:algo} we describe the compilation method that we use to find a quantum circuit implementing the Hamiltonian evolution.
This method does {\it not} exploit specific features of gauge theories and is applicable to any Hamiltonian.
The method is composed of many parts; all but the final one have been known in the literature. To our knowledge they have never been put together as we do here.

In Sec.~\ref{sec:examples} we apply this method to a couple of models to test our compilation method.
These models, an ensemble of random diagonal Hamiltonians and the $\mathbb{Z}_2$ gauge theory coupled to staggered fermions, have been studied before and quantum circuits for them are known.
We find that our method produces shorter circuits than those previously published.

In Sec.~\ref{sec:gauge-therory} we study two possible {qubitizations} of $SU(2)$ gauge theory, with a four- and a five-dimensional Hilbert space per link, respectively.
These models have the exact $SU(2)$ gauge invariance; thus, by standard universality arguments, one can suspect that they have the same continuum limit as $SU(2)$ gauge theory.
\replaced{One could reasonably conjecture}{It is hard to imagine} that a finite-dimensional model in the universality class of $SU(2)$ gauge theories \replaced{cannot}{can} be substantially simpler than these models and have a time evolution that can be approximated by a significantly smaller number of gates.
We find that, for the four-dimensional model, one time evolution step involves 180 {\small CNOT} gates per plaquette, and the five-dimensional model involves 17,168 {\small CNOT} gates. The large circuit depth shows the importance of the space field truncation and stresses the need for truncations that are in the same universal class as the continuum model.

Finally, in Sec.~\ref{sec:discussion} we summarize the results and speculate about some directions to pursue significantly reduced gate costs of simulating gauge theories.

\section{The compilation method}
\label{sec:algo}

We describe here a method to take a given finite-dimensional Hamiltonian and encode its time evolution into a series of quantum gates; in other words, we describe the quantum compiler method we used.
In order to facilitate use with most quantum programming languages, we compose our circuits only with common gates; in particular, we make heavy use of {\small CNOT} gates and one-qubit rotation gates, defined by
\begingroup
\allowdisplaybreaks
\begin{align}
    \mathrm{CNOT} &= \ket{0}\!\bra{0}\otimes I + \ket{1}\!\bra{1}\otimes X,
    \label{eq:cnot-gate} \\
    R_i(\phi) &\doteq e^{-\mathrm{i}\phi \sigma_i/2} = \cos\bigg(\frac{\phi}{2}\bigg)I - \mathrm{i}\sin\bigg(\frac{\phi}{2}\bigg)\sigma_i,
    \label{eq:rot-gate}
\end{align}
\endgroup
for $0\leq\phi\leq4\pi$, all expressed in the computational basis $\{\ket{0},\ket{1}\}$ with qubits. Here, we denote the Pauli gates $\{X,Y,Z\}\doteq\{\sigma_1,\sigma_2,\sigma_3\}$ and the one-qubit identity gate by $I$.
Notably, these rotations include the typical phase gate $S=R_z(\pi/2)$ (and its inverse $S^\dagger$) and easily generate the Hadamard gate $H=XR_y(\pi/2)$
(up to an irrelevant global phase); more broadly, we can perform arbitrary one-qubit unitary operations, sometimes denoted in an Euler angle basis $U3(\theta,\phi,\lambda)\equiv R_z(\phi)R_y(\theta)R_z(\lambda)$ up to a global phase.
We denote the control ($a$) and target ($b$) qubits for each {\small CNOT} gate using subscripts $\mathrm{CNOT}_{a,b}$, corresponding to the former and latter factors in Eq.~\eqref{eq:cnot-gate}, respectively.
For simplicity of the analysis in this work, we assume a hardware with all-to-all connectivity between the qubits\added{, and we do not include an error mitigation strategy in our resulting circuits.} The noisiest quantum gates on current hardware are the 
{\small CNOT} gates, so we aim to minimize their use.
In fact, the number of {\small CNOT} gates will be used as a measure of how good the algorithm is.

It is convenient to break down our compilation method into five overall steps.

{\bf 1) Trotterization:}
The first step is to break up the time evolution operator over a time $T$ into $\mathcal{N}$ terms $H=\sum_{i=1}^\mathcal{N} h_i$ with matrix norms $\lVert h_i \rVert\sim1$ over several time steps of duration $\delta t =T/N_t$:
\begin{equation}
    U(T) = e^{-\mathrm{i} HT} \approx \left(\prod_{i=1}^\mathcal{N} e^{-\mathrm{i} h_i\delta t}\right)^{N_t},
\end{equation}
incurring an overall error of $\mathcal{O}(\delta t)$, 
due to the nonzero commutators among $h_i$.
Also, we point out that, in the case of gauge theories, most actions are a sum over plaquettes (or slightly larger Wilson loops), which commute if they do not share a link.
So, one can then split the Hamiltonian into two commuting sets of terms, $E$(ven) and $O$(dd), and evolve each set independently:
\begin{align}
    e^{-i H\delta t} &= e^{-\mathrm{i} \sum_\Box H_\Box \delta t} \\
    &\approx e^{-\mathrm{i}\sum_{E}H_{E}\delta t} e^{-\mathrm{i} \sum_{O} H_{O} \delta t}, \nonumber
\end{align}
incurring in kind an error of $\mathcal{O}(\lVert [H_E,H_O] \rVert\delta t^2)$ for a time step $\delta t$.
Since the plaquettes within each set commute, their corresponding quantum gates can be applied in parallel.
Consequently, in the $SU(2)$ gauge theory examples we will consider later, it will suffice for us to determine circuits to simulate a single plaquette going forward.

{\bf 2) Decomposition into Pauli strings:}
Suppose we encode the degrees of freedom of one plaquette into $n$ qubits.
The Hilbert space on which $H$ acts is then $2^n$-dimensional.
Hermitian $H$ can be written as a real linear combination of Pauli strings on $n$ qubits:
\begin{align}\label{eq:paulistrings}
    H = \sum_{j=0}^{4^n-1} c_j P_j,
\end{align}
where 
$P_j\in\{I,X,Y,Z\}^{\otimes n}$. 
Further, the inner product on Pauli strings,
\begin{align}
    \frac{1}{2^n}\mathrm{tr}[P_iP_j] = \delta_{ij},
    \label{eq:inner-pauli}
\end{align}
allows one to easily calculate the coefficients above:
\begin{align}
    c_j = \frac{1}{2^n}\mathrm{tr}[P_jH].
\end{align}

{\bf 3) Separation of Pauli strings into commuting sets:}
Since commuting Pauli matrices can be simultaneously diagonalized, it is useful \added{to} split all Pauli strings into commuting sets or ``clusters,'' so that the quantum gates used in diagonalizing each set need to be applied only once.
The full time evolution can be Trotterized as the evolution given by each cluster, whose respective Trotterizations each incur no error due to commutativity.

It is possible to efficiently decide whether two Pauli strings commute, by counting the number of positions at which the Pauli matrices of the two strings do not commute. The two Pauli strings commute if and only if this number is even.
Moreover, in general, partitioning Pauli strings into clusters of commuting operators can be performed sequentially.
Namely, one can process the strings in the order they are given, placing each string in the first available cluster.
If no cluster is available, we create a new one and insert the string.
As a note of caution for general use, this clustering is not guaranteed to result in the minimal number of clusters,\footnote{As we will see later in Sec.~\ref{sec:examples}, one can reduce the {\small CNOT}-cost per Pauli string for a quantum circuit simulating a Hamiltonian with these Pauli strings if one packs each cluster to be as large as possible.}
and its outcome in principle can depends crucially on the order in which the strings are processed.
Therefore, we will reformulate the clustering in terms of the graph coloring problem, as previously described in Refs.~\cite{2019arXiv190707859J,2020JChPh.152l4114V,vandenBerg2020circuitoptimization,9248636}.

In this formulation, we construct a graph $\mathcal{G}$ whose vertices are the Pauli strings, where an edge exists between any two Pauli strings $P_i$ and $P_j$ ($j\neq i$) if they do \emph{not} commute.
We recall that a vertex coloring of $\mathcal{G}$ assigns colors to the vertices such that any two adjacent vertices have different colors ~\cite{graphtheorybook}.
Therefore, all Pauli strings with the same color will commute.
In this formalism, finding the optimal clustering is coloring $\mathcal{G}$ with the least required number of colors, known as the chromatic number of $\mathcal{G}$.  Finding the chromatic number of a graph is known to be NP-complete~\cite{Karp1972},
and so there is no known efficient algorithm to do so.
Nevertheless, several approximation algorithms provide a coloring close to optimal \cite{10.1145/359094.359101,doi:10.1137/0720013}.
In the examples we discuss in Sec.~\ref{sec:examples}, we were able to find convenient commuting clusters with relative ease.

{\bf 4) Diagonalization of each cluster:}
This step describes how to construct the quantum circuit of an operator that simultaneously diagonalizes the commuting Pauli strings.
The approach that we use entails representing the strings in the \textit{tableau} formalism described in Refs.~\cite{PhysRevA.70.052328,vandenBerg2020circuitoptimization,2019arXiv190713623G}.

In particular, we will follow the presentation in Ref.~\cite{vandenBerg2020circuitoptimization} where the tableau is composed of the $\mathcal{X}$, $\mathcal{Z}$, and $\mathcal{S}$ blocks, which we will define below.
Suppose we have $\mathcal{N}$ Pauli strings, each of length $n$. Then, the $\mathcal{X}$ and $\mathcal{Z}$ blocks are
$\mathcal{N}\times n$ matrices,
while the $\mathcal{S}$ block is
a $\mathcal{N}$-dimensional column vector.
For the $i$th Pauli string $P_i$ in a given ordering with coefficient $c_i \neq 0$, 
the entries to the corresponding row of each block are given by
\begin{equation}
    \begin{split}
    \mathcal{X}_{ij} &= \begin{cases}
    0 \text{ if the } j\text{th digit of } P_i \text{ is } Z \text{ or } I\\
    1 \text{ otherwise}
    \end{cases}\\
    \mathcal{Z}_{ij} &= \begin{cases}
    0 \text{ if the } j\text{th digit of } P_i \text{ is } X \text{ or } I\\
    1 \text{ otherwise}
    \end{cases}\\
    \mathcal{S}_i &= \begin{cases}
    0 \text{ if } c_i > 0\\
    1 \text{ otherwise}
    \end{cases}
    \end{split}
\end{equation}
In other words, $\mathcal{X}_{ij}$ and $\mathcal{Z}_{ij}$ encode whether the $j$th digit of the Pauli string $P_i$ contains a factor of $X$ and $Z$, respectively, while $\mathcal{S}_i$ simply encodes the sign of the Pauli string's coefficient. Note that a Pauli string $P_i$ is diagonal if and only if it contains only factors $I$ and $Z$, or, equivalently, all the entries in row $i$ of $\mathcal{X}$ are $0$.
\deleted{In general, this task can be completed with at most $\mathcal{O}(n^2)$ two-qubit gates}~\cite{vandenBerg2020circuitoptimization,Miller:2022sol}\deleted{, and there exist several algorithms to diagonalize these clusters though not necessarily with the minimal number of two-qubit gates}~\cite{crawford2019,vandenBerg2020circuitoptimization}
\deleted{.}
Consequently, the task of simultaneously diagonalizing commuting Pauli strings may be viewed as applying conjugations with unitary gates that reduce all entries of $\mathcal{X}$ to $0$.
Since we seek to transform a Pauli string to another, these unitary transformations are Clifford gates, generated by Hadamard, phase, and {\small CNOT} gates.
The conjugation of these gates acts on the tableau according to the following rules \cite{PhysRevA.70.052328}:
\begin{itemize}
    \item \textbf{Hadamard gate conjugation on qubit j} or $H\left(j\right)$:
    \begin{align}
        \begin{split}
        \label{eq:hadconj}
        \mathcal{S}_i &\leftarrow \mathcal{S}_i \oplus \mathcal{X}_{ij}\mathcal{Z}_{ij} \\
        (\mathcal{X}_{ij},\mathcal{Z}_{ij}) &\leftarrow (\mathcal{Z}_{ij},\mathcal{X}_{ij})
        \end{split}
    \end{align}
    for all $i \,= 1,\, \ldots,\, \mathcal{N}$. That is, we flip the sign of a Pauli string if the Pauli matrix at position $j$ is $Y$. Then, swap the $j$th columns of $\mathcal{X}$ and $\mathcal{Z}$; practically speaking, we replace any factors $X$ or $Z$ in the $j$th digit of each string with $Z$ or $X$ respectively.
    \item \textbf{Phase gate conjugation on qubit j} or $S\left(j\right)$: 
    \begin{align}
        \begin{split}
        \label{eq:phaseconj}
            \mathcal{S}_{i} &\leftarrow \mathcal{S}_{i} \oplus \mathcal{X}_{ij} \mathcal{Z}_{ij}\\
            \mathcal{Z}_{ij} &\leftarrow \mathcal{Z}_{ij} \oplus \mathcal{X}_{ij}
        \end{split}
    \end{align}
    for all $i \,= 1,\, \ldots,\, \mathcal{N}$. That is, for the Pauli string $P_i$ encoded in row $i$ we flip the sign of its coefficient if its $j$th digit is $Y$,
    and we replace any factors $X$ or $Y$ in the $j$th digit of each string with $Y$ or $X$ respectively.
    \item \textbf{{\small CNOT} conjugation with control qubit a and target b} or $\mathrm{CNOT}\left(a,b\right)$:
    \begin{align}
        \begin{split}
        \label{eq:cnotconj}
            \mathcal{S}_{i} &\leftarrow \mathcal{S}_{i} \oplus \mathcal{X}_{ia}\mathcal{Z}_{ib}\left(\mathcal{X}_{ib} \oplus \mathcal{Z}_{ia} \oplus 1\right)\\
            \mathcal{Z}_{ia} &\leftarrow \mathcal{Z}_{ia} \oplus \mathcal{Z}_{ib}\\
            \mathcal{X}_{ib} &\leftarrow \mathcal{X}_{ib} \oplus \mathcal{X}_{ia}
        \end{split}
    \end{align}
    for all $i \,= 1,\, \ldots,\, \mathcal{N}$, and the interpretation is similar to those in Eqs.~\eqref{eq:hadconj} and \eqref{eq:phaseconj}.
\end{itemize}
Note that the update on $\mathcal{S}$ should be applied first in each case because the sign update depends on the current values of $\mathcal{X}$ and $\mathcal{Z}$. Here, the symbol~$\oplus$ denotes the addition modulo 2.

\begin{figure}
\centering
\includegraphics[]{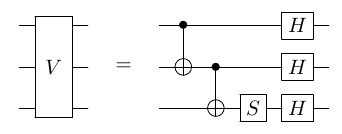}
\caption{Operator $V$ diagonalizing the Hamiltonian $H = IXX + ZYZ + XXI$}
\label{fig:ex-diagonalization}
\end{figure}

As an example, consider a Hamiltonian with mutually commuting Pauli strings $H = IXX + ZYZ + XXI$ \footnote{From now on, we omit the tensor product sign.}.
The $\mathcal{X}$, $\mathcal{Z}$, and $\mathcal{S}$ blocks are
\begin{align}
    \mathcal{X} &= \begin{pmatrix}
    0 & 1 & 1\\
    0 & 1 & 0\\
    1 & 1 & 0
    \end{pmatrix},
    &
    \mathcal{Z} &= \begin{pmatrix}
    0 & 0 & 0\\
    1 & 1 & 1\\
    0 & 0 & 0
    \end{pmatrix},
    &
    \mathcal{S} &= \begin{pmatrix}
    0\\
    0\\
    0
    \end{pmatrix},
\end{align}
labeling the qubits, left to right, from 1 to 3.
We can use $\mathrm{CNOT}_{1,2}$ and $\mathrm{CNOT}_{2,3}$ to nullify the first two columns of the $\mathcal{Z}$ stack.
Then, we observe that the third columns of the $\mathcal{X}$ and $\mathcal{Z}$ stacks are identical. Hence, a phase gate conjugation on the third qubit will nullify the $\mathcal{Z}$ stack. Finally, we apply Hadamard gates on all the three qubits to swap the columns of $\mathcal{X}$ and $\mathcal{Z}$. Thus, all of the resulting entries of the $\mathcal{X}$ stack are all 0. Meanwhile, the $\mathcal{Z}$ and $\mathcal{S}$ stacks become
\begin{align}
    \mathcal{Z} &= \begin{pmatrix}
    0 & 1 & 0\\
    0 & 1 & 1\\
    1 & 0 & 0
    \end{pmatrix}, &
    \mathcal{S} &= \begin{pmatrix}
    0\\
    1\\
    0
    \end{pmatrix},
\end{align}
respectively.
The resulting diagonal Pauli strings are together $V\,H\,V^\dagger = IZI - IZZ + ZII$, where we denote the diagonalizing circuit by $V$. Figure~\ref{fig:ex-diagonalization} depicts the circuit diagonalizing $H$.

\added{In general, it is possible to simultaneously diagonalize an arbitrary cluster of commuting Pauli strings with at most $\mathcal{O}(n^2)$ two-qubit gates}~\cite{vandenBerg2020circuitoptimization,Miller:2022sol}\added{. Moreover, there exist several algorithms to accomplish this diagonalization, although not necessarily with the minimal number of two qubit gates}~\cite{crawford2019,vandenBerg2020circuitoptimization}\added{.}

{\bf 5) Exponentiation of $I/Z$ Pauli strings via a Binary Tree:}
Finally, we describe our procedure to compile a quantum circuit simulating a Hamiltonian composed of $\mathcal{N}$ \emph{diagonal} Pauli strings (i.e., in $\{I,Z\}^{\otimes n}$).
In particular, we propose a bookkeeping process to search for shared {\small CNOT} gates between strings within the cluster, in contrast with existing methods, which compile a circuit individually reducing each Pauli string with {\small CNOT} conjugations (described below) and then look for cancellations \emph{post hoc}.

\begin{figure}[htbp]
\begin{center}
\subfloat[\label{fig:xyz-us}]{
\includegraphics[]{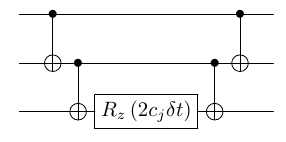}

}
\\
\subfloat[\label{fig:xyz-nc}]{
\includegraphics[scale=0.9]{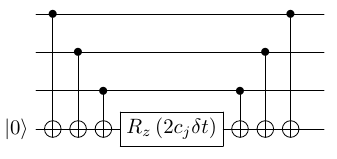}

 }
\end{center}
\caption{Quantum circuit implementing the factor $\exp(-\mathrm{i} c_j ZZZ \delta t)$.
(a) No ancillary qubit is used. 
Notably, there are other combinations of {\small CNOT} conjugations that one may perform to arrive at other equivalent circuits implementing this operator.
(b) One ancillary qubit, prepared as $\ket{0}$, is the target of each {\small CNOT} and the site of the rotation, as prescribed in this more common method (c.f. Ref.~\cite{Nielsen+Chuang}).
}
\label{fig:xyz}
\end{figure}

To start, for a single such Pauli string, we may use Eq.~\eqref{eq:cnotconj}---more specifically, the identity
\begin{equation}
    Z\otimes Z = {\text{CNOT}}_{12}\ (I \otimes Z)\ {\text{CNOT}_{12}}.
\end{equation}
By successive applications of this identity we can map any diagonal Pauli string to a string containing a single factor of $Z$.
Exponentiating such Pauli strings can then be accomplished with a single one-qubit rotation gate, $R_z(2 c_j \delta t)$ on a given qubit:
\begin{equation}
    e^{-\mathrm{i} c_j I \cdots I Z I\cdots I \delta t} = I \cdots I R_z(2 c_j \delta t) I\cdots I
\end{equation} 
This procedure is equivalent to the gate decomposition of Refs.~\cite{2021arXiv210903371L,Miller:2022sol}. 
For example, a circuit implementing $\exp(-\mathrm{i}c_j ZZZ \delta t)$ in such a fashion is shown in Fig.~\subref*{fig:xyz-us}.

This prescription deviates slightly from the usual technique (e.g., Ref.~\cite{Nielsen+Chuang}), which utilizes a single ancillary qubit on which the rotation is performed instead.
Via that alternative procedure, there are more {\small CNOT} conjugations required and fewer opportunities for the gates involved to cancel.
We display the same example, $ZZZ$, using this procedure in Fig.~\subref*{fig:xyz-nc}.


Equipped now with this generic strategy of reducing individual Pauli strings to one-qubit rotations, we describe in the remainder of this section
our own compilation method to efficiently simulate the diagonal Pauli strings together.
Our procedure seeks to induce sharing of {\small CNOT} gates within the cluster,
in the same spirit of reducing circuit depth as in step 4, where we share {\small CNOT} costs over a cluster via simultaneous diagonalization.
In particular, as we apply the first {\small CNOT} gate in each conjugation reducing a Pauli string to a rotation, we keep track of those conjugations in a list or ``stack'' and classically perform the {\small CNOT} conjugations on the remaining strings in the cluster, as per Eq.~\eqref{eq:cnotconj}. Moreover, we refrain from applying the second {\small CNOT} gate in the conjugation until after the rotation gates of all other Pauli strings have been executed on the circuit.
In essence, this approach can be understood as inserting the identity (i.e., a pair of identical, consecutive {\small CNOT} gates to cancel) between every pair of remaining Pauli strings in the cluster.
We provide instructions to efficiently carry out this procedure below, in terms of a tree traversal algorithm.
(See e.g., Ref.~\cite{goodrich2011data} for a review of tree data structures and traversal.)

\begin{figure}[htbp]
\centering
\includegraphics[]{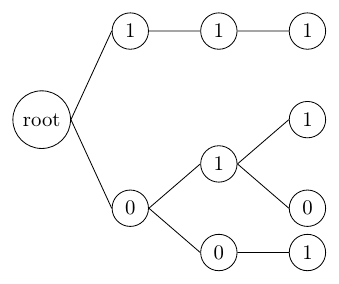}
\caption{Tree representation of $IIZ + IZI + IZZ + ZZZ$. The value inside each node along a branch indicates the gate $I$ or $Z$ on each qubit for the corresponding Pauli string term. For a general linear combination, the coefficient of a Pauli string is stored with its corresponding leaf.
}
\label{fig:tree-repr-example}
\end{figure}

We proceed by representing the linear combination of $\mathcal{N} \geq 1$ Pauli strings as a binary tree with $n$ levels where $n$ is the total number of qubits (equivalently the length of the Pauli strings).
One calls the zeroth level of the tree the ``root'' and the nodes of the $n$th and final level the ``leaves;'' notably the order of these $n$ tree levels may be a permutation of the corresponding qubit digits of the Pauli strings, equivalent to a temporary reordering of the qubit labels on our circuit.\footnote{For example, one may arrange the levels to label the qubits in descending order of number of Pauli strings sharing a factor $Z$ on that digit. We found this modified ordering of tree levels yielded modest savings $\mathcal{O}(n)$ in {\small CNOT} cost in some cases, when compared with the straightforward ordering where qubit $j$ is labeled by tree level $j$, and resulted in greater cost in other cases. \added{At this time, the cause of this higher cost is not clear. Determining how the {\small CNOT} cost depends on the relabeling of tree levels is the subject of further research.}} 
Each tree branch (i.e., path from the root to a leaf) corresponds to a Pauli string, and the values of its $n$ nodes are 0 or 1, corresponding to the factors $I$ or $Z$ respectively of the string.
Figure \ref{fig:tree-repr-example} shows the tree representation for the Hamiltonian $H = IIZ + IZI + IZZ + ZZZ$. For example, the values of the nodes in the bottom branch are $0$, $0$, and $1$, corresponding to the Pauli string $IIZ$.
For an arbitrary linear coefficient of Pauli strings, one includes the coefficient of Pauli string in the corresponding leaf node. (This task can be done for example by creating a leaf object that inherits all the properties of a node object but has an additional coefficient attribute.)


\begin{figure}[H]
\includegraphics[width=0.48\textwidth]{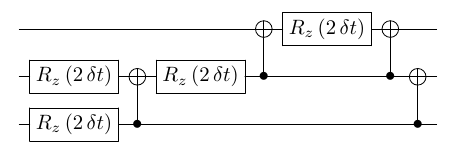}
\caption{Circuit simulating one Trotterized time step with the Hamiltonian $H = IIZ + IZI + IZZ + ZZZ$. Here $\delta t$ is the Trotterization time step. The qubits are ordered from top to bottom (i.e., as increasing levels in the tree structure of Fig.~\ref{fig:tree-repr-example}). }
\label{fig:ex-circuit}
\end{figure}

Mapping a commuting cluster of Pauli strings to this tree structure, we now present how to derive a quantum circuit from a traversal of this tree.
Our routine traverses the tree via a preorder Depth-First-Search (DFS) (see e.g., Ref.~\cite{goodrich2011data}).\footnote{We comment in passing that other orders are possible, equivalent to implementing Pauli strings in a different ordering.}
Reading along each branch from the root to its leaf, when a pair of nodes both with value 1 is encountered---say, at levels $i$ and $j>i$ --- a $\mathrm{CNOT}_{ji}$ gate will be applied on the circuit. 
We then classically perform a $\mathrm{CNOT}_{ji}$ conjugation to all Pauli strings on the tree, 
i.e., update the values along their branches according to Eq.~\eqref{eq:cnotconj}; the child node ($j$) value is flipped when the parent node ($i$) value is 1.
Lastly, the latter {\small CNOT} gate in the conjugation is stored in a stack, if there is no copy already in the stack; otherwise, we discard both the {\small CNOT} gate and its copy in the stack (an operation we will justify later). 
By this procedure, after reaching the leaf
there will be precisely one node ($i$) with value 1 along the branch, corresponding to a rotation on qubit $i$,
$R_z(2\,c\,\delta t)$, where $c$ is the coefficient of the string stored with the leaf/branch.
The procedure continues until the last branch is processed. Then, the {\small CNOT} gates from the stack are applied to the end of the circuit, from the latest to the earliest gate added into the stack.
Figure \ref{fig:ex-circuit} depicts the quantum circuit obtained for the example $H=IIZ+IZI+IZZ+ZZZ$.

Now, we explain why we may discard a {\small CNOT} gate and its copy in the stack.
First, we note that all {\small CNOT} gates with the same target qubit appear consecutively in the stack
and commute with each other.
Additionally, the control qubit $j$ for each $\mathrm{CNOT}_{ji}$ always corresponds to the child node in a pair $i,j$.
Therefore, 
the stack will have at most $\binom{n}{2}$ {\small CNOT} gates. 
There may be opportunities for further {\small CNOT} gate simplification particularly within this stack due to cancellations described in Refs.~\cite{2006quant.ph..4001M,2021arXiv210903371L}; however, by an argument listed below, we observe that such cancellations will not eliminate the bulk of the remaining {\small CNOT} cost in our final circuit.

To conclude, we comment on the cancellations among {\small CNOT} gates induced by this algorithm.
Within a cluster of commuting Pauli strings acting on $n$ qubits, we may have in general $\mathcal{N}\leq2^n$ such strings and in fact we typically have $\mathcal{N}\gg n$.
So, without any {\small CNOT} gate cancellation, a simulation of each Pauli string independently
could in principle require $\mathcal{O}(n\mathcal{N})$ {\small CNOT} gates on the circuit before all one-qubit rotations have been performed and equally as many after (i.e., in the stack).
However, on the stack there can only be $\mathcal{O}(n^2)$ such gates as we have argued above, so cancellation among {\small CNOT} gates in this stack alone can yield a factor of $\sim2$ savings in {\small CNOT} cost; the bulk of possible {\small CNOT} gates 
in the stack will cancel for large enough $\mathcal{N}\gg n$.
Moreover, we find that at least for 
our models in consideration, the {\small CNOT} cost turns out to be only $\mathcal{O}(\mathcal{N})$, due to other savings among the {\small CNOT} gates performed before one-qubit rotations.
In Appendix~\ref{sec:cnot-cost} we establish $\sim \mathcal{N}$ to be a generic lower bound on the number of {\small CNOT} gates entailed for
direct simulations of $\mathcal{N}$ (diagonal) Pauli strings.
We also stress that
the use of a tree is a data structure choice. For example, it is possible to store the Pauli strings in a list instead and perform the same {\small CNOT} conjugations according to our procedure. However, processing this list can result in more classical operations 
surpassing the cost of 
the binary tree. 
\\

The above discussion summarizes the method to decompose the time evolution into ordinary quantum gates.
It is important to note that steps 4 and 5, the most involved in the whole method, can be bypassed at the expense of arriving at an algorithm with a larger number of {\small CNOT} gates.
In fact, we can use {\small $H$} and {\small $S$} gates to diagonalize each of the Pauli strings separately; however, the increase in {\small CNOT} cost is substantial in the cases we considered.

\section{Examples}\label{sec:examples}

We now apply the general method outlined in the previous section to a couple of specific models. These examples -- random diagonal Hamiltonians and $\mathbb{Z}_2$ gauge theory -- are discussed here in order to establish the method as competitive with other methods one might consider. The $\mathbb{Z}_2$ gauge theory
is very simple and, actually, has been simulated in real quantum computers (with tiny lattices).



\subsection{Random Diagonal Hamiltonians}
\label{sec:random-hamiltonian}

To assess the relative performance of our method discussed particularly in step 5 above, we consider {\small CNOT} costs to simulate already diagonalized $n$-qubit Hamiltonians, which may be written as a linear combination of $\mathcal{N}\leq 2^n$ operators $\{I,Z\}^{\otimes n}$.

Consider first the  $\mathcal{N}= 2^n$ case.
It is known that the time evolution operator $U=\exp(-\mathrm{i}H\delta t)$ due to such a Hamiltonian can be implemented exactly with $\mathcal{O}\left(n\,2^n\right)$ {\small CNOT} gates because each Pauli string can be implemented with at most $(n-1)$ {\small CNOT} gates.
A previous study~\cite{2003quant.ph..3039B} showed that this unitary operator can be realized with $2^{n+1} - 3$ {\small CNOT} gates, roughly twice the minimal cost. 
In Appendix~\ref{sec:app-complete-hamiltonian}, we show that our method produces a circuit with $2^n - 2$ {\small CNOT} gates for arbitrary $n\geq2$ qubits, achieving the optimal cost derived in Ref.~\cite{2003quant.ph..3039B}. 
This result lends credence that the method we outlined is competitive with any other method currently available.

In the $\mathcal{N}< 2^n$ case, there are many choices of Hamiltonians and the final circuit  depends on the specific Pauli strings contained in the Hamiltonian. Thus, we applied our compilation method to a random sample of $M=10$ Hamiltonians with $\mathcal{N}$ diagonal Pauli strings and consider how the number of CNOT gates depends on $\mathcal{N}$.
There exist $\binom{2^n - 1}{\mathcal{N}}$ distinct sets of $\mathcal{N}$ Pauli strings, and so there are certain cases where the number of possible Hamiltonians is less than $M$; in these cases, we would have explored all the $<M$ possible Hamiltonians of $\mathcal{N}$ Pauli strings. Otherwise, the $M$ Pauli strings we have explored comprise only a small subset of the Hamiltonians with $\mathcal{N}$ Pauli strings. Of course, we expect that increasing the value of $M$ will produce yet more precise estimates of average {\small CNOT} cost. For each value of $\mathcal{N}$, we calculate a mean and standard error on the mean.

\begin{figure}[htbp]
    \centering
    \includegraphics[scale=0.59]{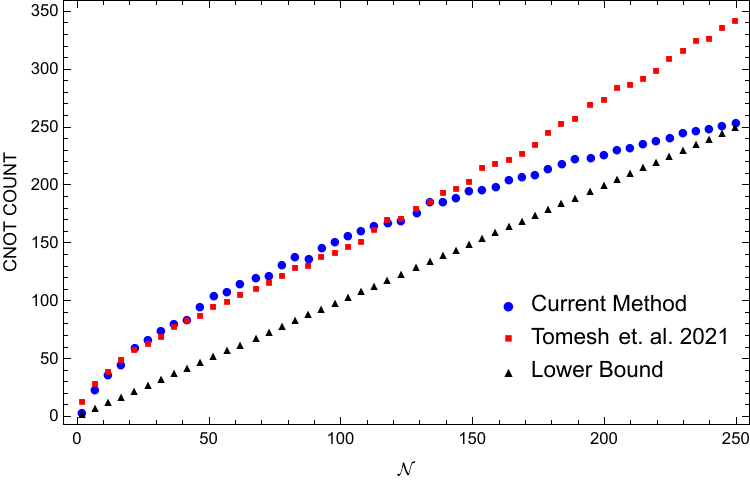}
    \caption{
    The average {\small CNOT} gate count required to simulate a diagonal Hamiltonian as function of the number of Pauli strings for $n = 8$ qubits. 
    The blue data points are obtained using the compilation method described in step 5. 
    The red data points are obtained using the method in~\cite{Tomesh:2021pns}. 
    A lower bound for each case is the number of Pauli strings acting nontrivially on more than one qubit, as derived in Appendix~\ref{sec:cnot-cost}.
    \added{Uncertainty bars derived from statistical sample sizes of 10 Pauli strings per choice of $\mathcal{N}$ are smaller than the data points depicted above.}
    }
    \label{fig:n8-us-vs-chicago}
\end{figure}

A recent study~\cite{Tomesh:2021pns} formulated the problem of minimizing the {\small CNOT} gates as solving the Traveling Salesman Problem (TSP) (see e.g., Ref.~\cite{Karp1972}). In this case, the vertices of the graph correspond to commuting Pauli strings and the weight on each edge is given by a {\small CNOT}-cost function they define. Given that the TSP is in general  NP-Complete~\cite{Karp1972}, they use the Christofides-Serdyukov (CS) algorithm, which returns a solution with a {\small CNOT} cost that is at most a factor of $1.5 \times$ the minimal cost~\cite{christofides1976worst,2020arXiv200402437V}. Moreover, they demonstrate that their ordering of Pauli strings on a quantum circuit based on the CS algorithm outperforms standard solutions such as those based on lexicographical ordering ~\cite{2018arXiv181202233T,2014arXiv1403.1539H}, ``Deplete Group,'' and ``Magnitude'' ordering ~\cite{2019Entrp..21.1218T}. We therefore apply the method in Ref.~\cite{Tomesh:2021pns} on the diagonal Pauli strings we generated. Figure~\ref{fig:n8-us-vs-chicago} shows the {\small CNOT} cost with the number of Pauli strings for $n = 8$ qubits.
We also include a lower bound derived in Appendix~\ref{sec:cnot-cost}.
The resulting {\small CNOT} gate counts are comparable for small $\mathcal{N}$.
However, as $\mathcal{N}$ increases, our procedure yields a smaller number of {\small CNOT} gates. In addition, the count from this current method approaches this lower bound while the other does not.



\subsection{\texorpdfstring{$\mathbb{Z}_2$}{Lg} Gauge Theory}
\label{sec:z2stag}

As a first example of  a gauge theory,
we consider the smallest nontrivial group, $\mathbb{Z}_2$, for our gauge symmetry.
Despite involving only a small discrete, abelian group,
its simplicity as well as its familiarity in previous literature (e.g., Refs.~\cite{PhysRevA.98.032331,Gustafson:2020yfe,Lamm:2019bik}) on quantum simulations of lattice gauge theories make it a natural candidate for introducing our methods.
Furthermore, the case of a $\mathbb{Z}_2$ gauge coupled to staggered fermions yields a slightly less trivial example, where comparison of our relatively low circuit depth with that of circuits suggested in past studies~\cite{Lamm:2019bik} will show an early benefit to our methods of circuit compilation.

In particular, we consider here staggered fermions in $1+1$ dimensions, as given in Ref.~\cite{Lamm:2019bik} and similar to a related model in Ref.~\cite{PhysRevA.98.032331}. There, an ancillary qubit was used to simulate each term of the Hamiltonian
\begin{align}
    \label{eq:h-z2}
    H =& \sum_{i=1}^L -\frac{m}{2}(-1)^i\sigma_3(i) + \sum_{i=1}^{L-1} \sigma_1(i,i+1)
    \nonumber \\
    &+ \frac{1}{4} \sum_{i=1}^{L-1} (-1)^i\sigma_3(i,i+1)[\sigma_1(i)\sigma_1(i+1)
    \nonumber \\
    &\phantom{ \frac{1}{4} \sum_{i=1}^L (-1)^i\sigma_3(i,i+1)[ }+\sigma_2(i)\sigma_2(i+1)],
\end{align}
where $L$ is the number of fermion sites, $m$ is the fermion mass, $\sigma_k(i)$ denotes Pauli matrices $\sigma_k$ applied to the subspace for the $i$th fermion site, and $\sigma_k(i,i+1)$ denotes $\sigma_k$ applied to the subspace for the gauge link connecting sites $i$ and $i+1$.

\begin{figure*}
\centering
\includegraphics[scale=0.9]{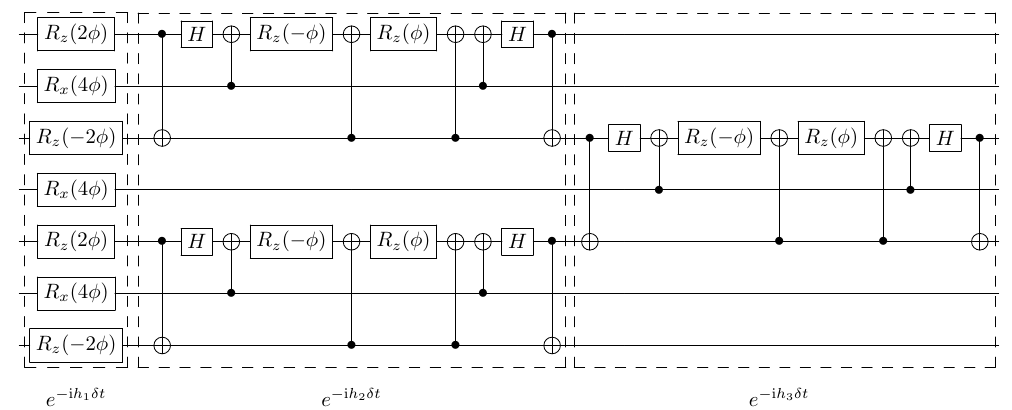}
\caption{
Quantum circuit approximately simulating one time step for a $\mathbb{Z}_2$ gauge theory with staggered fermions in $1+1$ dimensions for $L = 4$ lattice sites without a periodic boundary condition.
We use the first qubit to encode the first fermion site, the second qubit for the first gauge link, the third qubit for the second site, and so on.
The boxes denote the clusters of commuting Pauli strings, which were partitioned as follows:
the Hamiltonian $h_1$ consists of the one-body terms of Eq.~\eqref{eq:h-z2}, while $h_2$ consists of the multi-site terms acting on the gauge links $(1,2)$ and $(3,4)$, and $h_3$ likewise consists of the multi-site terms acting on link $(2,3)$.
We define the phase $\phi\equiv\Delta t/2$.
}
\label{fig:qc-z2}
\end{figure*}

Choosing \replaced{$L=4$ }{$V=4$} with open boundary conditions, an encoding of the global Hilbert space with only seven qubits is sufficient.
However, with an extra eighth, ancillary qubit included in the methods of Ref.~\cite{Lamm:2019bik}, there are fewer opportunities for rearrangement and cancellation of {\small CNOT} gates on the quantum circuit.
Consequently, the total {\small CNOT} cost for simulation of one time step was found to be 36.
Using the methods of section Sec.~\ref{sec:algo}, one can perform the same simulation without this ancilla and therefore reduce a {\small CNOT} cost to only 18.
The Hamiltonian in Eq.~\eqref{eq:h-z2} is a linear combination of 13 Pauli strings.
Using the sequential algorithm described in step 3 in Sec.~\ref{sec:algo}, we obtain three clusters of commuting Pauli strings, containing
seven, four, and two Pauli strings.
Each cluster was diagonalized as discussed in Sec.~\ref{sec:algo} before implementing our routine to construct the quantum circuit for each of the corresponding diagonalized clusters of Pauli strings.  Setting the fermion mass $m = 1$, the whole quantum circuit is shown in Fig.~\ref{fig:qc-z2}.

This benefit for a relatively simple model shows promise that our methods for circuit compilation will help keep {\small CNOT} cost relatively low compared to standard methods as we proceed to consider larger gauge groups.

\section{Qubitization of Non-Abelian Gauge Theories}
\label{sec:gauge-therory}

We will now consider a broader class of models where the gauge dynamical degrees of freedom can be thought as lying on the links of a spatial cubic lattice. The Hilbert space is the product of the $\ell$ local Hilbert spaces of each link (on $q$ qubits): $\mathcal{H} = (\mathbb{C}^{2^q})^{\otimes \ell}$ with a Hamiltonian of the form
\begin{align}\label{eq:plaquette}
    H &= g^2 K - \frac{1}{g^2} \sum_{x,\mu,\nu}{\rm tr}\left( U_{x,\mu} U_{x+\mu, \nu} U_{x+\nu,\mu}^\dagger U_{x,\nu}^\dagger \right),
\end{align} where $U_{x,\mu}$ is an operator acting on the Hilbert space of the link connecting sites $x$ and $x+\mu$\footnote{Here, we use the notation used in lattice gauge theory where the nearest point to $x$ in the direction $\mu$ is denoted by $x+\mu$.}.
The ``kinetic" term $K$ will be specified later as it depends on the model considered.
In a standard formulation of quantum \added{$SU(N)$} gauge theories, each entry of the \replaced{$N\times N$}{$2\times2$} unitary matrix \added{$U_{x,\mu}$} is \added{itself} an operator acting on an infinite-dimensional Hilbert space.
Here, instead, \added{we take} each entry \replaced{to act}{acts} on a finite-dimensional \replaced{operator. Thus}{space; thus}, \added{for $SU(2)$ gauge theory, on which we will focus the remainder of this section, we take}
$U = \openone\otimes \Gamma_4 + \mathrm{i} \sigma_k\otimes \Gamma_k$\added{,} where the matrices $\Gamma_\alpha$ ($\alpha=1,2,3,4$) will depend on the specific model.
Notice that for each $x$ and $\mu$, $U_{x,\mu}$ carries two pairs of matrix indices: one taking values $1,2$ and another taking values up to the dimension of~$\Gamma_\alpha$.

\replaced{Furthermore, a}{A} local $SU(2)$ gauge transformation corresponds to an assignment of a  different element of $SU(2)$ at each site.
Under gauge transformations the link operators $U$ transform as
\begin{equation}
    U_{ab} \mapsto L_{aa'} U_{a'b'} (R^\dagger)_{b'b},
\end{equation}
with $a,b=1,2$ and two elements of $SU(2)$, $L = \exp(\mathrm{i} \alpha_k \sigma_k/2)$ and $R = \exp(\mathrm{i} \beta_k \sigma_k/2)$.
These transformations
clearly leave the plaquette term in \eq{eq:plaquette} unchanged.
They are implemented within the Hilbert space as
\begin{equation}\label{eq:hilbert-gauge}
    U_{ab} \mapsto  L_{aa'} \,U_{a'b'} \,(R^\dagger)_{b'b}
    = \mathcal{R} \, \mathcal{L} \, U \, \mathcal{L}^\dagger \, \mathcal{R}^\dagger,
\end{equation}
with $\mathcal{L}=\exp(\mathrm{i} \alpha_k J^L_k)$, $\mathcal{R}=\exp(\mathrm{i} \alpha_k J^R_k)$
for certain matrices $J^L_k, J^R_k$ ($k=1,2,3$) with the same dimension as $\Gamma_\alpha$. In Ref.~\cite{Orland:1989st} it was shown that the 10  matrices $\Gamma_\alpha, J^L_k, J^R_k$ satisfy \eq{eq:hilbert-gauge} if they are chosen to be the generators of a representation of $\mathrm{SO}(5)$.
A\added{s such, a} choice of irreducible representation (irrep) \added{of $\mathrm{SO}(5)$} gives a model with a finite-dimensional Hilbert space and exact $SU(2)$ gauge symmetry and, therefore, a candidate for a qubitization of $SU(2)$ gauge theory. Generically, the smaller irreps yield lower-dimensional Hilbert spaces for the theory, and fewer quantum gates are expected to represent its time evolution. The two smallest irreps are discussed here.

\subsection{\texorpdfstring{$N=4$}{Lg} (spinor) irrep}
\label{sec:spinnor-irrep}

The smallest irrep of $\mathrm{SO}(5)$ is the four-dimensional spinor representation given by
\begin{eqnarray}
  \Gamma_4 &=& \dfrac{1}{2}\begin{pmatrix} 0 & 1\\1 & 0 \end{pmatrix}, \quad
\Gamma_k = \dfrac{1}{2} \begin{pmatrix} 0 & -\mathrm{i}\sigma_k\\\mathrm{i}\sigma_k & 0 \end{pmatrix}, \nonumber\\
J_k^L &=& \dfrac{1}{2} \begin{pmatrix} \sigma_k & 0 \\ 0 & 0 \end{pmatrix}, \quad
J_k^R = \dfrac{1}{2} \begin{pmatrix} 0 & 0 \\ 0 &  \sigma_k \end{pmatrix}.
\end{eqnarray}
However, this list of operators so far only suggests the form of a plaquette term, while $J_k^L J_k^L + J_k^R J_k^R$ is proportional to the identity.
A theory with only a single plaquette term cannot have a continuum limit, because the eigenvectors of the Hamiltonian are independent of the only parameter in the theory, $g^2$. Therefore, the correlation length of the ground state cannot be made arbitrarily large (in units of lattice spacing). As such, it is necessary to include another term to make eigenvectors and, consequently, correlation lengths tunable.
In the four-dimensional representation, there is one term
\begin{equation}
    K = \sum_l \Gamma_5(l),
\end{equation}
where
\begin{equation}
   \Gamma_5  = \frac{1}{2}\begin{pmatrix} I & 0\\0 & -I \end{pmatrix}.
\end{equation}
Together, the plaquette term and $K$ form a plausible qubitization of the $SU(2)$ gauge theory.
This model was first considered in Ref.~\cite{Orland:1989st} where it was speculated to be in the same universality class as $SU(2)$ gauge theory in 3+1 spacetime dimensions.
In order to encode this formulation on a quantum computer, only two qubits per link are required.

Now, we discuss the results of applying our method to this model. The $K$ term consists of four Pauli strings, which are all diagonal one-qubit operators; they can all be implemented without any {\small CNOT} gate. \added{Next, we turn to the plaquette term in Eq.~\eqref{eq:plaquette}. After tracing out the shared two-dimensional space, we can write it as $ \sum_{i,j,k,l = 1}^4\,c_{ijkl}\,\Gamma_{ijkl}$ where $c_{ijkl}\equiv\mathrm{tr}[\sigma_i\sigma_j\sigma_k\sigma_l]$ (taking $\sigma_4\equiv I$) and $\Gamma_{ijkl} \equiv \Gamma_i\otimes \Gamma_j\otimes \Gamma_k \otimes \Gamma_l$. By a counting argument or explicit computation, one finds that only 64 of the coefficients $c_{ijkl}$ are non-zero. In addition, each $\Gamma_k$ in this representation corresponds to a single Pauli string. Therefore, the expansion of the plaquette term results in $64$ Pauli strings.}


Using the sequential algorithm described in step 3, these latter $64$ operators can be grouped into four clusters, each containing $16$ Pauli strings.\deleted{Diagonalizing each cluster required no more than nine {\small CNOT} gates, and} We used $2 \times 28 = 56$ {\small CNOT} gates altogether to diagonalize these clusters.
With each cluster diagonalized, we can apply the method developed in step 5 to compile the time evolution circuit.\deleted{We observe that each cluster required no more than $36$ {\small CNOT} gates in this step, and } The total {\small CNOT} gate count to compile the time evolution of all the diagonal clusters is $124$.
Thus, we use $56 + 124 = 180$ {\small CNOT} gates per plaquette to simulate a time evolution step of this model.

\subsection{\texorpdfstring{$N=5$}{Lg} (fundamental) irrep}

In a five-dimensional truncation, we consider the fundamental representation of $\mathrm{SO}(5)$, given by
\begingroup
\allowdisplaybreaks
\begin{eqnarray}
   \Gamma_1 =
\begin{pmatrix}
0&1&0&0&0\\
1&0&0&0&0\\
0&0&0&0&0\\
0&0&0&0&0\\
0&0&0&0&0
\end{pmatrix},\quad
\Gamma_2 &=&
\begin{pmatrix}
0&0&1&0&0\\
0&0&0&0&0\\
1&0&0&0&0\\
0&0&0&0&0\\
0&0&0&0&0
\end{pmatrix},\nonumber \\
\Gamma_3 =
\begin{pmatrix}
0&0&0&1&0\\
0&0&0&0&0\\
0&0&0&0&0\\
1&0&0&0&0\\
0&0&0&0&0
\end{pmatrix}, \quad
\Gamma_4 &=&
\begin{pmatrix}
0&0&0&0&1\\
0&0&0&0&0\\
0&0&0&0&0\\
0&0&0&0&0\\
1&0&0&0&0
\end{pmatrix}, \nonumber
\end{eqnarray}
\endgroup
\begingroup
\allowdisplaybreaks
\begin{eqnarray}
J^L_1 &=&
\dfrac{1}{2}\begin{pmatrix}
0&0&0&0&0\\
0&0&0&0&-i\\
0&0&0&-i&0\\
0&0&i&0&0\\
0&i&0&0&0
\end{pmatrix}, \nonumber \\
J^L_2 &=&
\dfrac{1}{2}\begin{pmatrix}
0&0&0&0&0\\
0&0&0&i&0\\
0&0&0&0&-i\\
0&-i&0&0&0\\
0&0&i&0&0
\end{pmatrix},\nonumber\\
J^L_3 &=&
\dfrac{1}{2}\begin{pmatrix}
0&0&0&0&0\\
0&0&-i&0&0\\
0&i&0&0&0\\
0&0&0&0&-i\\
0&0&0&i&0
\end{pmatrix},\nonumber\\
J^R_1 &=&
\dfrac{1}{2}\begin{pmatrix}
0&0&0&0&0\\
0&0&0&0&i\\
0&0&0&-i&0\\
0&0&i&0&0\\
0&-i&0&0&0
\end{pmatrix}, \nonumber\\
J^R_2 &=&
\dfrac{1}{2}\begin{pmatrix}
0&0&0&0&0\\
0&0&0&i&0\\
0&0&0&0&i\\
0&-i&0&0&0\\
0&0&-i&0&0
\end{pmatrix},\nonumber\\
J^R_3 &=&
\dfrac{1}{2}\begin{pmatrix}
0&0&0&0&0\\
0&0&-i&0&0\\
0&i&0&0&0\\
0&0&0&0&i\\
0&0&0&-i&0
\end{pmatrix}.
\end{eqnarray}
\endgroup
Here, a non-trivial ``kinetic term" of the form
\begin{equation}
    K = \sum_l K(l) = \sum_{l} J^L_k J^L_k(l)
\end{equation}
($l$ indexes the links) is possible; note here $J^L_k J^L_k=J^R_k J^R_k$.
This model is sometimes called the Horn model~\cite{HORN1981149}.\footnote{ \added{There has been work to generalize the Horn model by including the coupling to gradually larger representations of $SU(N)$ for $N\geq2$; however, while this effort may help extrapolate away the error of this truncation of the links~\cite{Byrnes:2005qx,Zohar:2014qma}, it necessarily relinquishes the relatively low Hilbert space dimension of the Horn model. } }

Three qubits are required to encode the five-dimensional Hilbert space in each link. Since $2^3=8=5+3$, three directions are ``wasted" in this encoding. The matrix elements of the Hamiltonian involving these extra three dimensions are unphysical and can be chosen arbitrarily, with no change in the Physics. Unfortunately, we have not been able to exploit
this freedom to simplify the quantum circuits.
Notably, one may attempt to encode the global Hilbert space at once in a smaller basis of $\lceil4\log_2(5)\rceil=10$ qubits, where a given qubit may actually include information from the local Hilbert spaces of multiple links.
However, though one might expect to reduce circuit depth by reducing the number of qubits in the encoding, we instead find that such an encoding results in an even greater circuit depth, as it requires more terms in the Hamiltonian in order to derive an operator that is applied only on the desired gauge link.

\begin{figure}[tbp]
\includegraphics[width=0.47\textwidth]{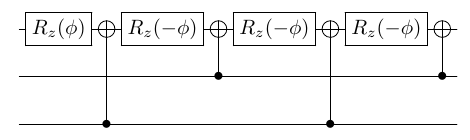}
\caption{
Quantum circuit simulating a time step for the kinetic term K as encoded in Eq.~\eqref{eq:K_l}. Here, we have set $\phi \equiv 3\delta t/8$.
}
\label{fig:kinetic}
\end{figure}

The operator $K$ can be decomposed into Pauli strings as (a sum over links of)
\begin{align}
    K(l) &= 
    \frac{3}{4}\,\mathrm{diag}
    (0,1,1,1,1,0,0,0)
    \label{eq:K_l} \\
    &= \frac{3}{16}(2III+ZII-ZIZ-ZZI-ZZZ), \nonumber
\end{align}
where we have appended each $K(l)$ with zero eigenvalues to encode over three qubits.
We note in passing that other paddings of these operators are possible, but we did not find significant changes to gate counts.
The operator $\exp(\mathrm{i}\phi III)$ corresponds to a global phase, having no effect on measurements, and is ignored.
Because these Pauli strings are already diagonal, we can immediately apply the routine discussed in the step 5 to obtain the quantum circuit in \fig{fig:kinetic}.
Note that this circuit has to be applied to every link individually. Therefore, the cost of applying $K$ for an elementary plaquette is 16 {\small CNOT} gates.

\replaced{As discussed in Sec.~\ref{sec:spinnor-irrep}, there are 64 terms in the $\sum_{i,j,k,l = 1}^4\,c_{ijkl}\,\Gamma_{ijkl}$ expansion of the plaquette term. In this representation, however, each $\Gamma_i$ can be decomposed into four Pauli strings.}{Now, we turn to the plaquette term in Eq.~\eqref{eq:plaquette}.
After tracing out the shared two-dimensional space, we can write it as $ \sum_{i,j,k,l = 1}^4\,c_{ijkl}\,\Gamma_{ijkl}$ where $c_{ijkl}\equiv\mathrm{tr}[\sigma_i\sigma_j\sigma_k\sigma_l]$ (taking $\sigma_4\equiv I$) and $\Gamma_{ijkl} \equiv \Gamma_i\otimes \Gamma_j\otimes \Gamma_k \otimes \Gamma_l$.
By a counting argument or explicit computation, one finds that only 64 of the coefficients $c_{ijkl}$ are non-zero. Furthermore, each encoded $\Gamma_i$ is decomposed into four Pauli strings.} Consequently, each $\Gamma_{ijkl}$ decomposes into $4 \times 4 \times 4 \times 4 = 256 $ Pauli strings, and each plaquette therefore is a combination of $64 \times 256 = 16,384$ Pauli strings.
By the sequential algorithm discussed in step 3 of the method, we find that these Pauli strings can be grouped into 64 clusters, each containing 256 commuting Pauli strings.
We diagonalize each cluster as per step 4 and compile the circuit of each resulting diagonal cluster as per step 5.

As expected, most of the complexity of this procedure lies in compiling the circuit of the diagonal cluster. \replaced{In fact, the total number of {\small CNOT} gates required for diagonalizing all the clusters is $2 \times 320 = 640$.}{In fact, each cluster was diagonalized with at most eight {\small CNOT} gates. The total number of {\small CNOT} gates required for diagonalizing all the clusters is 320. Since the diagonalizing circuit has to be uncomputed, we required 640 {\small CNOT} gates in cluster diagonalization.} \replaced{On the other hand, the total number of {\small CNOT} gates needed to compile all 64 diagonal clusters is 16,512.}{On the other hand, each diagonal cluster used at most 260 {\small CNOT} gates or approximately one {\small CNOT} gate per Pauli string. The total number of {\small CNOT} gates needed to compile all 64 diagonal clusters is 16,512 {\small CNOT} gates.} Therefore, we use 17,152 {\small CNOT} gates to compile one plaquette term of the Hamiltonian. Since 16 {\small CNOT} gates were used to compile the kinetic part of the Hamiltonian, we obtain a circuit with 17168 {\small CNOT} gates, or $\approx 1.05$ {\small CNOT} gates per Pauli string.

It is worth mentioning a recent proposal to exploit the symmetries of wave functions satisfying the Gauss's law condition of $SU(2)$ lattice gauge theory to more sparsely (i.e., in fewer dimensions) encode the nontrivial representations of $SU(2)\times SU(2)$ on qubits~\cite{Klco:2019evd,Banerjee:2017tjn}.
In particular, the space of wave functions for lattices whose vertices each see exactly three links meeting can be constrained at the classical level (i.e., before encoding on qubits) in a relatively simple fashion to solve the $SU(2)$ Gauss's law condition of the theory automatically and thereby reduce the $\mathbf{2}\oplus\mathbf{\bar{2}}$ representation subspace to one physical dimension.
This design results in a model requiring exactly one qubit per gauge link, however severely restricting the geometry of gauge links and vertices.
In order to avoid this restriction and more broadly allow for arbitrary $d$ spatial dimensions, we consider a square lattice geometry and focus our attention to the aforementioned encoding of each gauge link on qubits separately without yet imposing Gauss's law, ensuring a more modular nature to this procedure upon which we can build larger lattice model simulations.
Nonetheless, the methods we present may also be applied to a Hamiltonian with a different geometry or encoding, and we will return to a discussion of non-cubic lattice geometries later in Sec.~\ref{sec:discussion}.

\section{Discussion}\label{sec:discussion}
\replaced{
In this paper, we have offered a contribution to the very thorny and important problem of understanding, in detail, the resource requirements for quantum simulations---including but not limited to $SU(2)$ gauge theories. 
In particular, we assembled state-of-the-art quantum circuit compilation methods and showed specifically how $SU(2)$ gauge theory simulations would be prohibitive on digital quantum computers in the current era. 
Moreover, we can see from these methods that larger-dimensional gauge links for these theories will also necessitate quantum error correction codes to address the severe noise introduced by such large resource estimates. 
We hope that other efforts will build upon this work as a foundation, either to develop more efficient resource usage or perhaps to show these circuits to be as efficient as we can obtain. 
}{
The main message contained in our results is that the circuit depth entailed in quantum simulation of these $SU(2)$ gauge theories for arbitrary wave functions is prohibitively large in the current era of digital quantum computers. 
This disappointing result is not a rigorous lower bound, so it is important to keep in mind some directions of study that may circumvent this cost.
}

We believe that the compilation method we used
is at least competitive with the
other methods currently available, as it performs well in several cases where other methods have been tried or rigorous bounds are known.
Nevertheless, the compilation method is very generic, as it applies to any finite-dimensional Hamiltonian and makes no use of specific features of gauge theories.

In particular, the fact that the \emph{physical} Hilbert space is a small subspace of the theory (as many states are just gauge copies of each other) was not exploited here.
Conversely, it is not obvious how to utilize gauge invariance in this formulation, since it is generally acknowledged that rewriting a gauge theory in terms of gauge-invariant quantities leads to non-localities, which could make the circuit depths even larger.
Still, this reduction is a possible avenue for improvement, and perhaps a gauge-fixed version of the formalism can also be used for this purpose.

There is another sense in which the large Hilbert spaces we encountered could be \replaced{shrunk}{shrunken}.
For instance, the five-dimensional local Hilbert space of the Horn model was encoded onto three qubits, leaving $2^3-5=3$ dimensions unused. This redundancy is present in any qubitization with a local Hilbert space whose dimension is not a power of 2.
It is possible that some alternative encoding of these links will permit
a quantum circuit to properly evolve the five-dimensional subspace that actually describes each link of the model and yet has a different action on the remaining three dimensions from what we have prescribed.
At this time, we do not have a systematic way of exploiting this freedom to shorten circuits.

Finally, a key point: the large circuit depths we found stresses the importance of having a field space truncation with small dimensions. In particular, this point reinforces the need for qubitization in the same universality class as the continuum model.
This way, the final continuum limit will require an increase in the number of spatial lattice sites (linearly increasing the number of qubits required yet not the circuit depth) but not that the field truncation be lifted---a step that would enormously increase the circuit depth as we saw in the examples shown here.

\begin{acknowledgments}
This material is based upon work supported in part by the U.S. Department of Energy, Office of Nuclear Physics under Award Number(s) DE-SC0021143,  and DE‐FG02‐93ER40762, and DE-FG02-95ER40907.
The modules \textit{Qiskit}~\cite{Qiskit} and \textit{Numpy}~\cite{harris2020array} have been significantly used throughout this project.
\end{acknowledgments}

\appendix


\section{{\small CNOT}-cost lower bound of simulating Pauli strings}
\label{sec:cnot-cost}

In this appendix, we elaborate on the $\mathcal{O}(\mathcal{N})$ lower bound estimate of {\small CNOT} cost in simulating a diagonal Hamiltonian (See Sec.~\ref{sec:random-hamiltonian}), where $\mathcal{N}$ is the number of Pauli strings in its expansion.
This lower bound applies to methods that utilize {\small CNOT} conjugations to reduce a Pauli string into an one-qubit operator, as described at the beginning of step 5 in Sec.~\ref{sec:algo}.
We argue that at least one {\small CNOT} is required between the rotation gates for simulating diagonal $\exp(\mathrm{i}\phi_iP_i)$ and $\exp(\mathrm{i}\phi_jP_j)$ sequentially, for $j\neq i$ and with $P_i$ or $P_j$ consisting of at least two nontrivial factors. 
Here, the only restriction on the phases $\phi_i$ and $\phi_j$ is that they cannot both be equal to odd integer multiples of $\pi$.

Since $P_i$ and $P_j$ are diagonal, they may contain only factors $I$ (trivial) and $Z$ (non-trivial).
Let $\omega\left[P_i\right]$ be the number of non-trivial factors in $P_i$, or equivalently the Hamming weight of
$P_i$ (i.e., $\omega[I]=0$ and $\omega[Z]=1$).
In addition, let $C_i$ be a product of {\small CNOT} gates whose conjugation of $P_i$ reduces to a string of Hamming weight 1.
We will prove that $C_i\neq C_j$, implying the existence of at least one {\small CNOT} gate that may not be canceled when implementing $P_i$ and $P_j$ in sequence.

We can readily observe that $C_i \neq C_j$ if $\omega\left[P_i\right] \neq \omega\left[P_j\right]$, as this assumption implies that $C_i$ and $C_j$ are sequences of different lengths.
Therefore, we turn our focus to the case $\omega\left[P_i\right] = \omega\left[P_j\right]=k\geq2$, for which we will prove $C_i\neq C_j$ by induction on $k$.

First, consider the base case $k=2$:
Let $P_i\neq P_j$ and $\omega\left[P_i\right] = \omega\left[P_j\right] = 2$.
Perform a $\mathrm{CNOT}_{a,b}$ gate conjugation, with $a$ and $b$ chosen such that $P_i\mapsto P^\prime_i$ with $\omega\left[P^\prime_i\right] = 1$.
Suppose the same {\small CNOT} conjugation also results in $\omega\left[P^\prime_j\right] = 1$. Then Eq.~\eqref{eq:cnotconj} implies necessarily $P_i=P_j$, contradicting the assumption $P_i \neq P_j$.
(To see this conclusion, repeat the same conjugation again to $P^\prime_i$ and $P^\prime_j$, reproducing $P_i$ and $P_j$ respectively, which both have Hamming weight 2. Consequently, we see that $P_i^\prime$ and $P_j^\prime$ must each act nontrivially only on qubit $b$, meaning $P^\prime_i=P^\prime_j$.)
Therefore, there is a distinct {\small CNOT} conjugation needed to implement $P_j$, and hence $C_i\neq C_j$.

Having taken care of the base case, we state the induction hypothesis: there exists an integer $k \geq 2$ such that if $P_i\neq P_j$ while $\omega\left[P_i\right] = \omega\left[P_j\right] = k $ then $C_i\neq C_j$.
Finally, we prove the inductive step; consider the case where $P_i\neq P_j$ while $\omega\left[P_i\right] = \omega\left[P_j\right] = k + 1 $.
Perform a $\mathrm{CNOT}_{a,b}$ conjugation resulting in $P^\prime_i$ and $P^\prime_j$ such that $\omega[P^\prime_i]=k$.
If $\omega[P^\prime_j]\neq k$, then we are done, by the observation made before the beginning of this induction proof.
Now, if $\omega[P^\prime_j]=\omega[P^\prime_i]=k$, we consider the remaining cases: $P^\prime_i=P^\prime_j$ (already yielding a contradiction to our assumption $P_i\neq P_j$ by a similar reasoning as in the base case) and $P^\prime_i\neq P^\prime_j$.
In this latter case, we may invoke the inductive hypothesis, to see that $C_i'$ and $C_j'$ corresponding to $P_i'$ and $P_j'$ are not equal. Therefore, we also have $C_i\neq C_j$, completing our inductive step.



Iterating this argument over all consecutive pairs of the $\mathcal{N}$ diagonal Pauli strings in a given order for Trotterized time evolution, we find that there must be greater than $\mathcal{N}_2$ {\small CNOT} gates used to implement the complete circuit, where $\mathcal{N}_2$ is the number of Pauli strings with Hamming weight greater than one, which means $\mathcal{N}-(n+1)\leq \mathcal{N}_2\leq \mathcal{N}$.
In addition, when considering our method prescribed in Sec.~\ref{sec:algo}, there is a cost of uncomputing the {\small CNOT} conjugations instantiated with each $C_i$; however, we argue in step 5 of our procedure that this cost is $\mathcal{O}(n^2)$.
In kind, the diagonalization of Pauli strings has a {\small CNOT} cost of $\mathcal{O}(n^2)$ as well.
For $\mathcal{N}\gg n$, we thus expect a scaling in {\small CNOT} cost that is at least $\sim \mathcal{N}$.

It is worth noting that relative {\small CNOT} cost of simulating consecutive Pauli strings with one ancilla was found to define a metric space over these strings~\cite{Tomesh:2021pns}, which would imply a generic lower bound cost of at least $\sim \mathcal{N}$ as well.
However, we do not find this formulation for assessing relative {\small CNOT} costs to be applicable to our method.

\section{{\small CNOT} gate count for a diagonal Hamiltonian with \texorpdfstring{$\mathcal{N} = 2^n$}{Lg} Pauli Strings}
\label{sec:app-complete-hamiltonian}

In this section, we prove for general $n\geq2$ that, given a diagonal Hamiltonian with all $2^n - 1$ Pauli strings besides the identity, our algorithm realizes the exact time evolution with precisely $2^n - 2$ {\small CNOT} gates.
To simplify the wording of our proof, 
we also include the identity Pauli string, so that we have $2^n$ Pauli strings, contributing only an extra global phase and without changing the {\small CNOT} gate count.

First, recall our tree traversal procedure for compiling the circuit of a diagonal Hamiltonian in step 5 of Sec.~\ref{sec:algo}. We perform a {\small CNOT} gate conjugation once we encounter a node of value $1$ that has a parent with value $1$, flipping the value of the child node in the process.
For the tree representing a Hamiltonian composed of $2^n$ Pauli strings, every node has an equal number of children with values $0$ and $1$ at each deeper level.
Consequently, any {\small CNOT} gate conjugation preserves the number of children nodes of value $1$ that have a parent with value $1$, even if we perform the {\small CNOT} conjugation on multiple branches simultaneously.
Since a {\small CNOT} conjugation is applied when we encounter such a child node, the number of {\small CNOT} conjugations is precisely the number of such children.
With these observations in hand, we are ready to find the {\small CNOT} cost of a circuit prescribed by our procedure.

Neglecting the root of the tree, we label the levels by $l = 1, \ldots, n$. The number of level $l$ nodes that have a parent node with value $1$ is $2^{l-1} - 1$ in a complete binary tree (a binary tree in which every node besides the leaves has $2$ children).
Then, the total number of such nodes is $\sum_{l=1}^{n} 2^{l-1} - 1 = 2^n - \left(n + 1\right)$, which corresponds to the total number of {\small CNOT} conjugations.
As discussed in step 5, this quantity is precisely the number of {\small CNOT} gates that are implemented on the circuit before inserting any of the {\small CNOT} gates from the stack into the circuit.
Thus, determining the overall {\small CNOT} cost has been reduced to calculating the remaining number of {\small CNOT} gates in the stack.
Having every possible diagonal Pauli string in our Hamiltonian, we must perform every possible $\mathrm{CNOT}_{ji}$ with $j > i$ as prescribed.
For each $i,j$, the number of $\mathrm{CNOT}_{ji}$ gates in the stack is odd if and only if $j - i = 1$ because every node with value $1$ at a level $i$ has precisely $2^{\left(j - i\right) -1}$ children at level $j>i$ with value $1$.
Having the same target qubit, these {\small CNOT} gates are all consecutive in the stack, and we may cancel all the {\small CNOT} gates with $j - i \neq 1$, leaving only one of each {\small CNOT} gate with $j - i = 1$, of which there are $n  - 1$.
Hence, the total {\small CNOT} cost is therefore $2^n - (n + 1) + (n - 1) = 2 ^n - 2$.

\bibliography{references}

\end{document}